\documentclass[article]{elsarticle}

\usepackage{fixltx2e}
\usepackage{hyperref}
\usepackage{ dsfont }
\usepackage{amsmath }
\usepackage{lineno,hyperref}
\usepackage{ bbold }
\usepackage{ amssymb }
\usepackage{epstopdf}
\usepackage{cleveref}
\biboptions{numbers,sort&compress}
\modulolinenumbers[5]
\usepackage{color}

\usepackage{setspace}
\begin{document}

\begin{frontmatter}

\title{Dissipative solitons in forced cyclic and symmetric structures}

\author[myfirstaddress]{F. Fontanela\textsuperscript{*}}
\ead{filipe.fontanela15@imperial.ac.uk}
\author[mysecondaddress]{A. Grolet}
\author[myfirstaddress]{L. Salles}
\author[myfourthaddress]{A. Chabchoub}
\author[myfifthaddress]{A.R. Champneys}
\author[mysixthaddress]{S. Patsias}
\author[myfirstaddress,mysevenththaddress]{N. Hoffmann}

\address[myfirstaddress]{Department of Mechanical Engineering, Imperial College London, Exhibition Road, \\ SW7 2AZ London, UK}
\address[mysecondaddress]{Department of Mechanical Engineering, Arts et M\'etiers ParisTech, 8 Boulevard Louis XIV, 59000 Lille, France}
\address[myfourthaddress]{School of Civil Engineering, The University of Sydney, Sydney NSW 2006, Australia}
\address[myfifthaddress]{Department of Engineering Mathematics, University of Bristol, BS8 1TR Bristol, UK}
\address[mysixthaddress]{Rolls-Royce plc, PO Box 31, DE24 8BJ, Derby, UK}
\address[mysevenththaddress]{Department of Mechanical Engineering, Hamburg University of Technology, 21073 Hamburg, Germany}

\begin{abstract}
	
The emergence of localised vibrations in cyclic and symmetric rotating structures, such as bladed disks of aircraft engines, has challenged engineers in the past few decades. In the linear regime, localised states may arise due to a lack of symmetry, as for example induced by inhomogeneities. However, when structures deviate from the linear behaviour, e.g. due to material nonlinearities, geometric nonlinearities like large deformations, or other nonlinear elements like joints or friction interfaces, localised states may arise even in perfectly symmetric structures. In this paper, a system consisting of coupled Duffing oscillators with linear viscous damping is subjected to external travelling wave forcing. The system may be considered a minimal model for bladed disks in turbomachinery operating in the nonlinear regime, where such excitation may arise due to imbalance or aerodynamic excitation. We demonstrate that near the resonance, in this non-conservative regime, localised vibration states bifurcate from the travelling waves. Complex bifurcation diagrams result, comprising stable and unstable dissipative solitons. The localised solutions can also be continued numerically to a conservative limit, where solitons bifurcate from the backbone curves of the travelling waves at finite amplitudes. 

\end{abstract}

\begin{keyword}
Travelling wave excitation \sep solitons \sep vibration localisation \sep cyclic structures
\end{keyword}

\end{frontmatter}


\section{Introduction}

Localisation of vibrations in cyclic and symmetric structures is an important topic for rotating machines due to high cycle fatigue. In the case of linear vibrations, localisation may occur due to a lack of symmetry, induced e.g. by manufacturing variabilities or wear \cite{Hodges1982,Hodges1983,Bendiksen1987}. This problem, usually referred to as a mistuning problem, has attracted much attention in the literature, encompassing areas such as efficient numerical tools for analysis \cite{Castanier2006}, experimental investigations \cite{Judge2001}, and even the use of intentional mistuning to minimise localised vibrations \cite{Castanier2002}.

In real engineering applications, the assumption of linear vibrations has to be seen as an approximation, and the dynamics of complex engineering structures is usually a result of multiple linear and nonlinear phenomena. In the case of bladed disks of aircraft engines, nonlinearities may arise e.g. due to friction induced by internal joints \cite{Krack2016, Pesaresi2017}, or due to large deformations induced by strong excitations \cite{Grolet2012}. In the nonlinear regime, the processes which lead to localised vibrations may go beyond mistuning. It is well-known, for example, that even perfect cyclic and symmetric structures may localise vibrations due to bifurcations \cite{Vakakis1992,Vakakis1993,Vakakis1993b,King1995,Vakakis1996,Georgiades2009}. Moreover, the effects of nonlinearities in nonhomogeneous systems is still a very active topic of research not only in structural dynamics \cite{Capiez-Lernout2015}, but also in other research fields \cite{Ikeda2013, Ivanchenko2011,Bodyfelt2011}. 

This paper focuses on the nonlinear dynamics of cyclic and symmetric structures that are excited by forces that have the form of travelling waves. In the case of bladed disks, such excitations, also known as engine orders, are caused by aerodynamic or structural forces that move along the structure preserving their shape. Analysis is based on a minimal model composed of symmetric Duffing oscillators, cyclically connected to each other, and in the presence of linear velocity-proportional dampers. It has already been shown that, in the conservative regime, the underlying model has localised states composed of envelope solitons \cite{Grolet2016,Fontanela2017}. These results were obtained from analytical solutions of a Nonlinear Schr\"{o}dinger Equation (NLSE). For the non-conservative case, here a modified NLSE is developed, and analytical solutions describing localised states are not possible anymore. Therefore, general solutions and their stability are assessed numerically. Within this regime, it is demonstrated that localised states, sometimes called dissipative solitons \cite{Grelu2012,Akhmediev2008}, branch off from the travelling waves, resulting in complex bifurcation diagrams composed of stable and unstable single and multi-soliton states. Finally, these localised states are continued into the conservative limit and it is demonstrated that they become nonlinear normal modes of the underlying system. These modes are forced to bifurcate from the backbone curve of the travelling wave at finite vibration amplitude.

The paper is organized as follows. In Sec. \ref{Sec:MinMod} the minimal model for weakly nonlinear cyclic and symmetric structures is introduced. Moreover, the assumptions necessary to derive a non-conservative NLSE are discussed, and strategies to compute localised solutions and check their stability are presented. General numerical results are shown in Sec. \ref{Sec:NR}. We discuss how localised states branch off from travelling waves in the NLSE framework, and these results are verified through time integration of the full equations. In Sec. \ref{Sec:Cons}, localised states obtained from the non-conservative NLSE are continued into the conservative limit. Bifurcation diagrams show that localised states detach from travelling waves at non-zero amplitudes, contrary to the standard analytical predictions obtained from the conservative NLSE solved in an infinite spatial domain. Finally, Sec. \ref{Sec:Conc} discusses the main conclusions and suggests directions for further investigations. 

\section{The minimal model and solution methods} \label{Sec:MinMod}
	
\begin{figure}[h]
\begin{center}
	\vspace{-0.1cm}
	\includegraphics[trim=0.25cm 4.25cm 0.25cm 4.cm, clip=true, angle=0, scale=0.375]{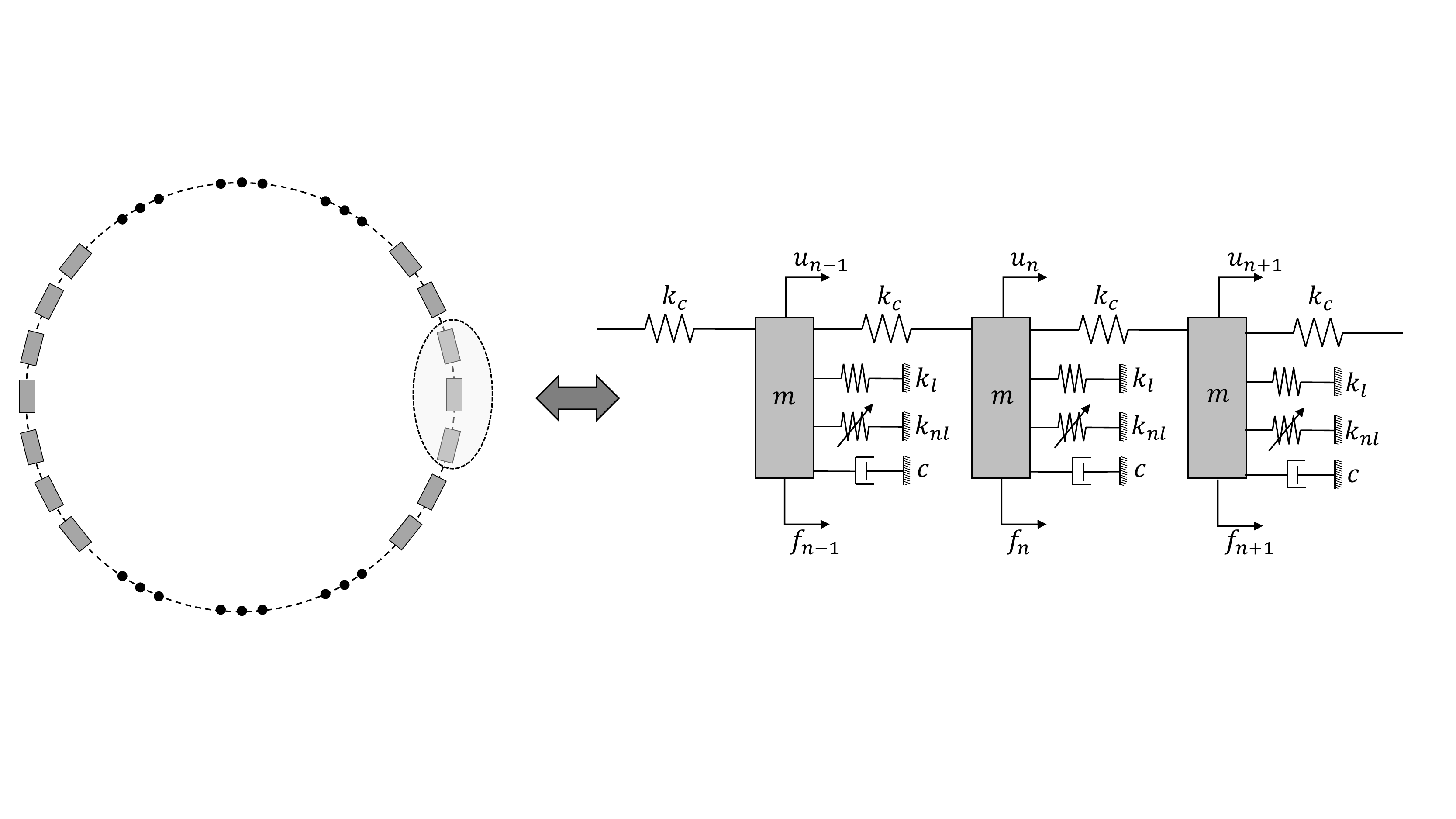}
	\caption{Physical system under investigation. In the left-hand side, an illustration of the full system, while the right-hand side displays any three neighbouring degrees of freedom.}
	\label{Fig:System}
\end{center}
\end{figure}

The physical system under investigation is displayed in Fig. \ref{Fig:System}. 
It consists of $N_s$ blocks of mass $m$, cyclically connected through linear springs $k_c$, and attached to the ground by linear and cubic springs $k_l$ and $k_{nl}$, respectively. The dissipation is modelled by linear viscous dampers $c$, connected to the ground, while the external force applied to the $n$th mass is $f_n$. The system in Fig. \ref{Fig:System} can be understood as a minimal model of a bladed disk \cite{Grolet2012,Grolet2011} operating in the strong deformation regime, while a positive value of $k_{nl}$ models the induced hardening effect.

The displacement of the $n$th degree of freedom is given by
\begin{equation}
m \ddot{u}_n + c\dot{u}_n + k_l u_n -k_c (u_{n-1}+u_{n+1} - 2 u_n) + k_{nl} u_n^3=f_n.
\end{equation}
This equation can be written in rescaled variables, for convenience, as 
\begin{equation}
\ddot{u}_n + \gamma^2 \dot{u}_n + \omega_0^2  u_n -\omega_c^2 (u_{n-1}+u_{n+1}-2u_n) + \xi u_n^3=g_n,
\label{Eq:Mot}
\end{equation}
where $\gamma^2=c/m$, $\omega_0^2=k_l/m$, $\omega_c^2= k_c/m$, $\xi=k_{nl}/m$ and $g_n=f_n/m$. In this paper, the investigation focuses on responses to travelling wave excitation. The system dynamics induced by these forces is a typical analysis in bladed disks vibrations. Travelling wave excitation is induced in rotating machines e.g. due to aerodynamic forces that move along the structure \cite{Petrov2004} or due to engine orders \cite{Jones2003}. Mathematically, a travelling wave excitation can be written as 
\begin{equation}
g_n(t)=F_0 \exp\{\mbox{i}\left[k(n - 1)a - \omega_f t \right] \} + \mbox{c.c.},
\label{Eq:TWforce}
\end{equation}
where $F_0$ is the force amplitude, $k$ is its wave number, $a=2\pi/N_s$ is the lattice parameter, $\omega_f$ is the external frequency, $\mbox{i}$ is the imaginary unity, and $\mbox{c.c.}$ represents the complex-conjugate of the preceding expression. It should be noted that the wave number $k$ can only assume integer numbers and, consequently, the external force respects the cyclic symmetry ($g_{N_s+1}=g_1$). 

\subsection{Linear analysis of the conservative system}

In the linear regime ($\xi=0$) and in the absence of damping ($\gamma^2=0$) and external forces ($F_0=0$), the system depicted in Fig. \ref{Fig:System} has exact travelling wave solutions of the form
\begin{equation}
u_n(t)=U_0 \exp\{\mbox{i}[k(n-1)a - \omega_k t]\} + \mbox{c.c.}. 
\label{Eq:TW}
\end{equation}
In Eq.\eqref{Eq:TW} the constant $U_0$ stands for the travelling wave amplitude, while $\omega_k=\sqrt{\omega_0^2 + 4\omega_c^2 \sin^2 \left( \frac{ka}{2}\right)}$ is its corresponding natural frequency. Again, the travelling wave solution is only valid when $k$ is an integer and, therefore, the response fulfils the cyclic symmetry ($u_{N_s+1}=u_1$).

\subsection{The slowly-varying and weakly nonlinear regime}

In the presence of damping ($\gamma^2 \ne 0$) and external forces ($F_0\ne 0$), this paper focuses on the slowly-varying weakly nonlinear regime of Eq.~\eqref{Eq:Mot}. Thus, the system response $u_n$ is written as 
\begin{equation}
u_n(t)=\epsilon \Psi(X,T)\exp\{\mbox{i}[k(n-1)a - \omega_k t] \} +  \mbox{c.c.},
\label{Eq:Mod}
\end{equation}
where $\Psi(X,T)$ is an envelope function which modulates the travelling wave, and $\epsilon$ is a small parameter. In Eq.~\eqref{Eq:Mod}, $X$ and $T$ are continuous variables such as $X=\epsilon x=\epsilon (n-1)a$ and $T=\epsilon t$. In the undamped case and in the absence of external forces, it is well-known that the evolution of the envelope function $\Psi$ can be approximated by a conservative NLSE \cite{Grolet2016,Remoissenet1994}. Mathematically, such an equation is obtained after substituting Eq.~\eqref{Eq:Mod} into Eq.~\eqref{Eq:Mot}, rejecting higher-order harmonics, and applying a continuum approximation. This assumption works well when the envelope function $\Psi$ varies much more slowly in space and time when compared to the carrier wave period ($2\pi/\omega_k$) and the lattice parameter ($2\pi/N_s$), respectively. For the non-conservative case, a similar approach can be applied if the following additional assumptions are considered: (1) the damping value $\gamma^2$ and {\color{black} the newly defined  detuning parameter $\delta_\omega = \omega_k - \omega_f$} are at order $\epsilon^2$; and (2) the external force level is at order $\epsilon^3$. As result, a similar equation for the evolution of $\Psi$ is obtained such as 
\begin{equation}
\mbox{i} \frac{\partial \Psi}{\partial \tau} + P \frac{\partial^2 \Psi}{\partial \eta^2} + Q |\Psi|^2\Psi = -\mbox{i} \Gamma \Psi - h \exp\{\mbox{i}\delta_\omega\tau\},
\label{Eq:AcNLS}
\end{equation}
where $\tau=\epsilon T$, while $\eta=X - c_g t$ is a frame moving with the group velocity $c_g=\frac{\mbox{d}\omega_k}{\mbox{d}k}$. In Eq~\eqref{Eq:AcNLS}, the parameter $P=\frac{1}{2}\frac{\mbox{d}^2\omega_k}{\mbox{d}k^2}$ accounts for the dispersion, $Q=-\frac{3\xi}{2\omega_k}$ is the nonlinear coefficient, $\Gamma=\frac{\gamma^2}{2}$ models the linear damping effect, and $h=\frac{F_0}{2\omega_k}$ is the external force. The derivation of Eq.~\eqref{Eq:AcNLS} is presented in the Appendix. For the undamped and unforced case, the right-hand side of Eq.~\eqref{Eq:AcNLS} is null and the system recovers the conservative NLSE. Within this regime, the NLSE is an integrable system and localised solutions, consisting of envelope solitons, can be obtained analytically (see e.g. Ref. \cite{Remoissenet1994}). For the forced and damped case Eq.~\eqref{Eq:AcNLS} loses integrability and localised solutions written analytically are not available in the literature. Therefore, solitons of the damped and driven NLSE are studied numerically in the following.

\subsection{Solution approach} \label{Sub:Solution}

As usual it is assumed that the travelling response waves are vibrating with the same frequency as the external excitation such as
\begin{equation} 
\Psi(\tau,\eta)=\psi(\tau,\eta)\exp\{\mbox{i} \delta_\omega \tau \}.
\label{Eq:SolAss}
\end{equation}
After substituting Eq.~\eqref{Eq:SolAss} into Eq.~\eqref{Eq:AcNLS}, the process leads to the autonomous system
\begin{equation}
\mbox{i} \frac{\partial \psi}{\partial \tau} - \delta_\omega \psi + P \frac{\partial^2 \psi}{\partial \eta^2} + Q |\psi|^2\psi =  -\mbox{i} \Gamma \psi - h.
\label{Eq:FinalPsi}
\end{equation}
For the homogeneous ($\frac{\partial^2 \psi}{\partial \eta^2}=0$) and standing ($\frac{\partial \psi}{\partial \tau}=0$) case, solutions of Eq.~\eqref{Eq:FinalPsi} can still be obtained analytically. In the case of localised standing waves, it has been shown, numerically, that they detach from the homogeneous solutions through bifurcations \cite{Terrones1990,Barashenkov1996,Barashenkov1998}. {\color{black} In these previous publications dissipative solitons were computed using the external excitation as a continuation parameter. However, in vibration engineering the standard way to characterise mechanical structures relies on the behaviour of frequency response functions; i.e. keeping $h$ constant while $\delta_{\omega}$ is varied. Therefore, we solve Eq. \eqref{Eq:FinalPsi} using $\delta_{\omega}$ as the continuation parameter. 
	
Another difference from previous studies (see e.g. Refs. \cite{Barashenkov1996,Barashenkov1998}) is the way we have computed localised solutions as bifurcations from standing waves ($\frac{\partial \psi}{\partial \tau}=0)$ rather than as continuation from the undamped limit. We use the popular continuation code AUTO by first rewriting  Eq.~\eqref{Eq:FinalPsi} as a system of two first-order complex nonlinear differential equations and then solving a two-point boundary-value problem. The theoretical framework implemented in AUTO can be checked in Refs.~\cite{Doedel1998,ChKuSa:96} and references therein.}

\subsection{Stability analysis}

In this paper, the approach developed in Ref. \cite{Barashenkov1996} is used to assess the stability of stationary solutions $\psi_0(\eta)$ obtained with AUTO. Thus, the expression 
\begin{equation}
\psi(\eta,\tau)=\psi_0(\eta)+ \Delta\psi(\eta,\tau)
\end{equation}
is substituted into Eq.~\eqref{Eq:FinalPsi}, where  $\Delta \psi$ is a small perturbation. When the analysis is considered up to order $\Delta \psi$, which results in a linear investigation, the stability analysis yields 
\begin{equation}
J\left( \frac{\partial y}{\partial \tau} + \Gamma y \right)= H y.
\label{Eq:Stab1}
\end{equation}
In Eq.~\eqref{Eq:Stab1}, the perturbation $\Delta \psi$  is split into real and imaginary parts such as $y(\eta,\tau)=(\mbox{Re}\{\Delta \psi\} \ \ \mbox{Im}\{\Delta \psi\})^T$, while $J$ and $H$ are the two matrices depending on the real $\psi_R=\mbox{Re}\{\psi_0\}$ and imaginary $\psi_I=\mbox{Im}\{\psi_0\}$ parts of $\psi_0$ written as
\begin{gather}
J=\left( \begin{array}{cc}
0 & 1 \\
-1 & 0 \end{array} \right),\\
H=\left( \begin{array}{cc}
-P \frac{\partial^2}{\partial \eta^2} + \delta_\omega - Q(3 \psi_R^2 + \psi_I^2) & -2Q\psi_I\psi_R \\
-2Q\psi_I\psi_R & -P \frac{\partial^2}{\partial \eta^2} + \delta_\omega - Q(3 \psi_I^2 + \psi_R^2) \end{array} \right).
\label{Eq:Stab2}
\end{gather}
After separating the time and space variables via $y(\eta,\tau)=z(\eta)\exp\{\lambda \tau\}$, the linear stability problem becomes 
\begin{equation}
\mu J z(\eta)=H z(\eta),
\end{equation}
where $\mu=\lambda + \Gamma$. Therefore, if $\mbox{Re}\{\mu\}<\Gamma$ the solution is stable, otherwise it is unstable. For the special case of homogeneous solutions, stability can be addressed analytically. It has already been proven that flat solutions of Eq.~\eqref{Eq:FinalPsi} can be modulationally unstable \cite{Terrones1990,Barashenkov1996}. For the case of localised solutions, stability has to be investigated numerically. Therefore, the solution $\psi_0$ is discretized in space and the second-order derivative in Eq.~\eqref{Eq:Stab2} is calculated through finite differences. A detailed explanation about the implementation of this numerical procedure can be checked in Ref. \cite{Barashenkov1996}.

\section{Numerical results} \label{Sec:NR}

\begin{figure}[h]
	\begin{center}
		\vspace{-0.1cm}
		\includegraphics[trim=3.cm 0.25cm 0.25cm 0.cm, clip=true, angle=0, scale=0.285]{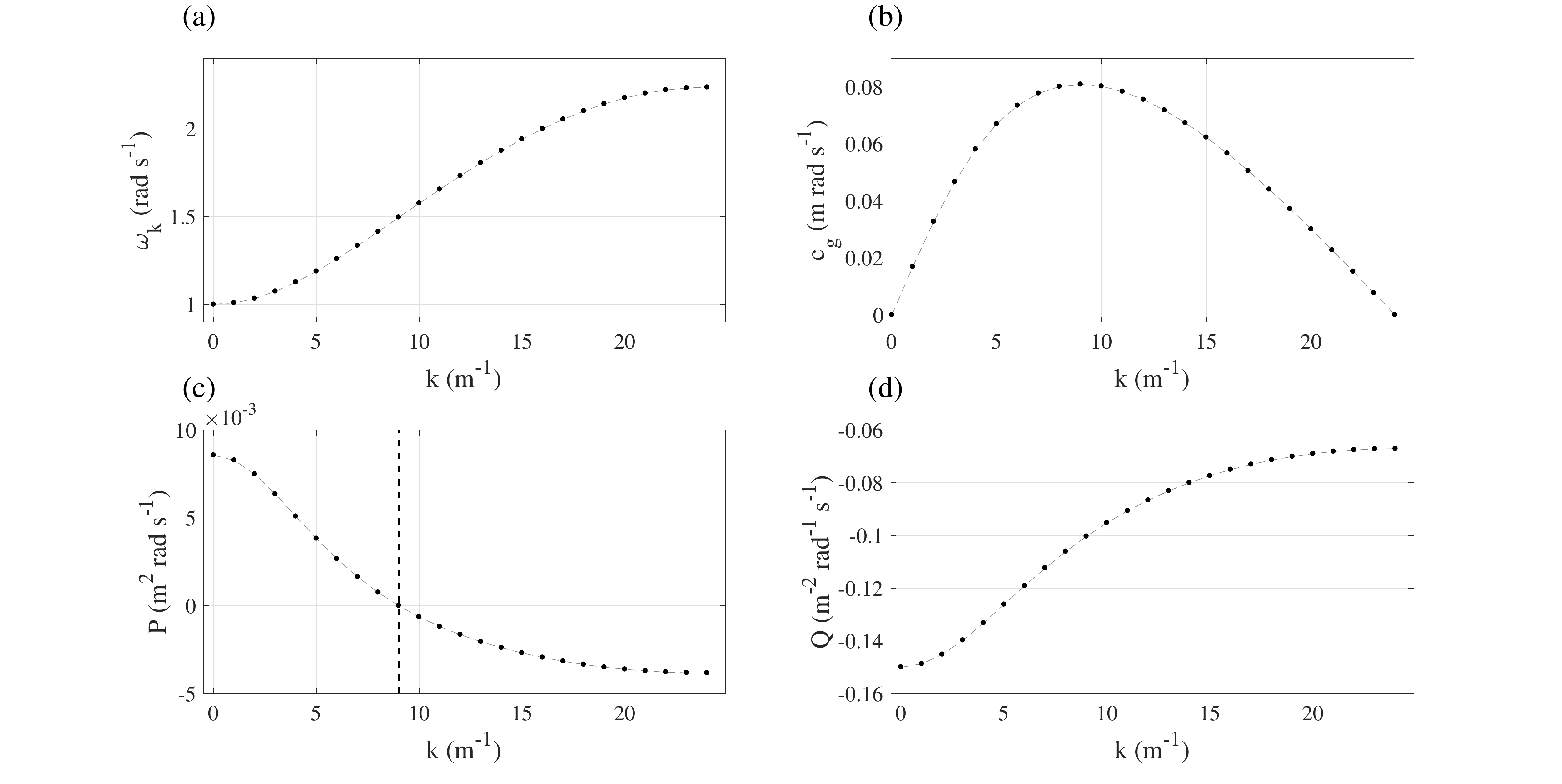}
		\caption{Evolution of the system parameters as a function of the wave number $k$: (a) natural frequency $\omega_k$; (b) group velocity $c_g$; (c) dispersion parameter $P$; and (d) nonlinear coefficient $Q$. The vertical dashed line in Panel (c) illustrates the transition from positive to negative values of $P$.}
		\label{Fig:SysPar}
	\end{center}
\end{figure}

The cyclic and symmetric system displayed in Fig. \ref{Fig:System} is investigated by assuming $N_s$=48 oscillators, $\omega_0^2$=1 s$^{-2}$, $\omega_c^2=1$ s$^{-2}$, $\xi=0.1$ m$^{-2}$s$^{-2}$, and $\gamma^2$=0.01 s$^{-1}$. Figure \ref{Fig:SysPar} displays the natural frequency $\omega_k$ of the linear system, the group velocity $c_g$, the dispersion parameter $P$, and the nonlinear coefficient $Q$, as functions of the wave number $k$. 
One should note that the system is cyclic and, therefore, the curves in Fig. \ref{Fig:SysPar} are symmetric with respect to $k$=24. Moreover, a transition from positive to negative values of $P$, at $k\approx$ 10 m$^{-1}$, is highlighted. This boundary separates two different regimes: for $PQ<0$ the system is in the defocusing range, while for $PQ>0$ the system is in the focusing range. It is well-known (see Refs. \cite{Remoissenet1994,Dauxois2006}) that plane waves in the focusing (defocusing) range and in an infinite domain are unstable (stable) when they are investigated in a conservative NLSE framework. Within the focusing range, solitons are stable solutions of the conservative system. In this section we investigate if soliton solutions also exist in the non-conservative regime. 

The strategy of solution proposed in Sub. \ref{Sub:Solution} is applied to Eq.~\eqref{Eq:FinalPsi}. Thus, standing solutions of the NLSE are investigated with AUTO. Due to the cyclic symmetry, any arbitrary solution $\psi_0(\eta)$ when translated in space is also a solution of Eq.~\eqref{Eq:FinalPsi}. In order to remove this symmetry in the continuation approach, Eq.~\eqref{Eq:FinalPsi} is solved for $\eta=[0, \pi]$ assuming Neumann boundary conditions ($\frac{\partial \psi}{\partial \eta}|_{\eta=0}$=$\frac{\partial \psi}{\partial \eta}|_{\eta=\pi}=0$). {\color{black} In this case, any general solution is forced to localise maximums and 
minimums of energy at the spatial boundaries $\eta=$ and $\eta=\pi$. 
Moreover, the solution for the remaining spatial configuration ($\eta=(\pi,2\pi)$) is obtained by simply reflecting the solution calculated 
for $\eta=[0,\pi]$. However, one should note that this implementation 
restricts the analysis to solutions which are symmetric with 
respect to $\eta=\pi$.}

{\color{black}
In the case of a single-hump soliton, the analysis should focus on the maximum 
amplitude of $\psi$. This value, when compared to the corresponding homogeneous states, identifies the level of localisation for an specific solution. However, the same measure is not sensitive to the number of humps in a
multiple-hump soliton. Therefore, a new measure 
\begin{equation}
E = \int_0^{2\pi} \left\lbrace |\psi|^2 + \left| \frac{\partial \psi}{\partial \eta}\right| ^2 + |\psi|^4 \right\rbrace \mbox{d}\eta 
\end{equation}
is introduced. One should note that $E$ is slightly different from the  energy of the envelope function used in Ref. \cite{Barashenkov1998}.}

\begin{figure}[]
	\begin{center}
		\vspace{-0.1cm}
		\includegraphics[trim=1.cm 0.25cm 0.25cm 1.cm, clip=true, angle=0, scale=0.2625]{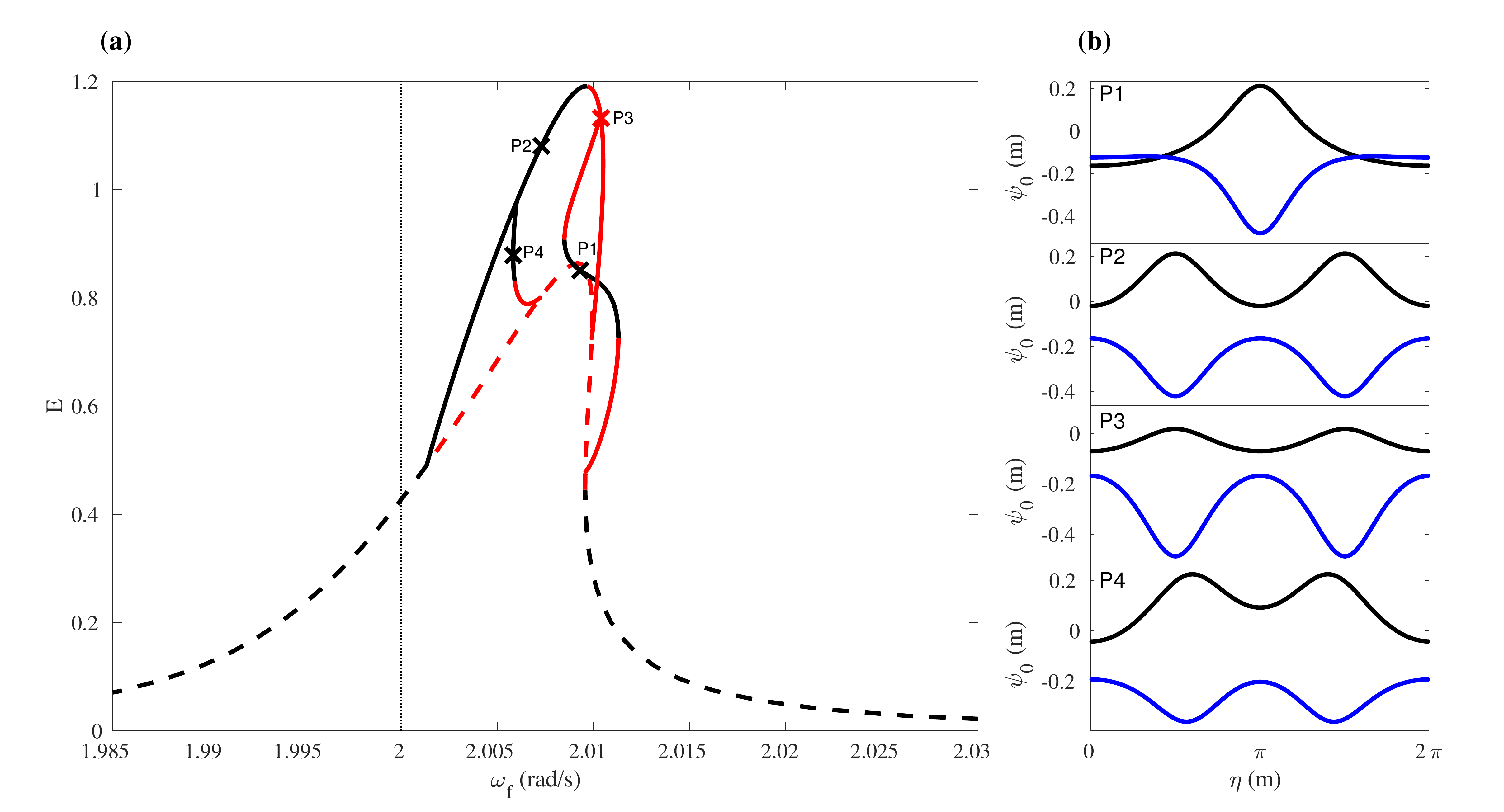}
		\caption{Solutions obtained from the NLSE implementation in Eq.~\eqref{Eq:FinalPsi} for $k$=16 m$^{-1}$ and $F_0$=0.007 ms$^{-2}$. Panel (a) illustrates $E$ as a function of $\omega_f$. The black dashed lines denote stable plane wave results, while the red one shows unstable plane wave solutions. The solid lines, which branch off from the travelling wave result, indicate stable (black) and unstable (red) localised solutions. The vertical dotted line shows the linear natural frequency of the system. Panel (b) shows the real (black) and imaginary (blue) parts of $\Psi_0$ at P1, P2, P3, and P4. }
		\label{Fig:h1}
		\vspace{0.5cm}
		\includegraphics[trim=1.cm 0.25cm 0.25cm .cm, clip=true, angle=0, scale=0.2625]{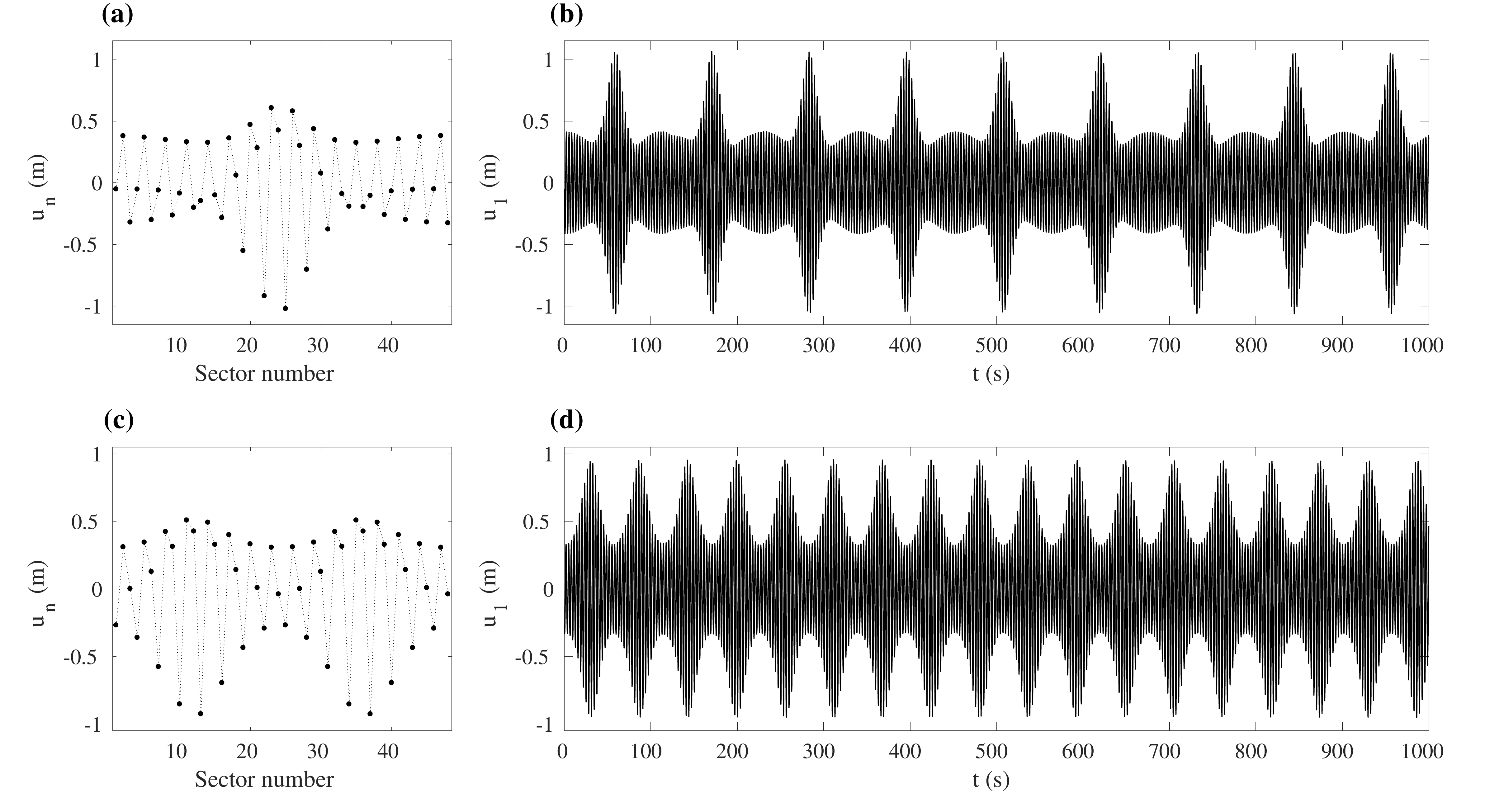}
		\caption{Results obtained through numerical time integration of the physical system for $k$=16 m$^{-1}$ and $F_0$=0.007 ms$^{-2}$. Panel (a): the spatial displacements at $t$=0 s calculated at P1 in Fig. \ref{Fig:h1}; Panel (b): time-evolution of the displacement $u_1(t)$ of the first oscillator. Panels (c) and (d) illustrate the same quantities for P2.}
		\label{Fig:t_h1}
	\end{center}
\end{figure}

Figure \ref{Fig:h1} displays $E$ as function of the external forcing frequency, calculated from the NLSE implementation.
The result is obtained by assuming the wave number $k$=16 m$^{-1}$ and the external excitation level $F_0$=0.007 ms$^{-2}$. This configuration leads the parameters of Eq.~\eqref{Eq:FinalPsi} to $P$=-0.00295 m$^{2}$rad s$^{-1}$, Q=-0.00750 m$^{-2}$rad$^{-1}$s$^{-1}$, $\Gamma$=0.005 s$^{-1}$ and $h$=0.00175 ms$^{-1}$. It is possible to verify, from Fig. \ref{Fig:h1}, that the plane waves loose stability at $\omega_f \approx$ 2.001 rad/s and localised solutions branch off from the travelling waves. These localised states consist of two hump solitons (see P2 and P3) which, as discussed before, are symmetric with respect to $\eta=\pi$ m. The branch detaches at $\omega_f \approx$ 2.001 rad/s with stable soliton solutions, but it looses stability after $\omega_f \approx $ 2.008 rad/s. The same analysis also shows two other branches of localised solutions: one detaching at $\omega_f \approx $ 2.008 rad/s; and another detaching at $\omega_f \approx$ 2.010 rad/s. 

An important feature, observed in the bifurcation diagram of Fig. \ref{Fig:h1}, is that localised solutions with one hump detaching from travelling waves do not necessarily continue to homogeneous states keeping only one hump as solutions. In fact, they can internally bifurcate from one hump to two humps (see points P1 and P2 in Fig. \ref{Fig:h1}). This behaviour resembles snaking bifurcations \cite{Hunt2000, Papangelo2017, Papangelo2018}, in which a branch of localised solutions fold to generate states with more or less localised elements, or a more or less widely extending zone of localised dynamics. In fact, a similar result has been already observed for solutions of Eq.~\eqref{Eq:FinalPsi}, but keeping the detuning parameter $\delta_\omega$ constant and varying the external force level $h$ \cite{Barashenkov1998}. 

In order to verify the predictions, initial states $u_n(0)$ and $\dot{u}_n(0)$ were computed from the NLSE implementation and the equations of the original physical system (Eq.~\eqref{Eq:Mot}) are calculated with a time-marching approach. Figure \ref{Fig:t_h1} illustrates the responses obtained assuming the stable states P1 and P2 of Fig. \ref{Fig:h1}. The initial displacement, in Panel (a), shows a localised hump on a travelling wave background. This hump keeps its shape while it moves around the system with roughly the group velocity, as shown by the displacement $u_1(t)$ in Panel (b). Panels (c) and (d) show similar results, but starting with a symmetric localised state composed of two humps. In this case, the displacement field $u_1(t)$ shows a doubly localised state moving around the system with a constant envelope profile. 

\begin{figure}[]
	\begin{center}
		\vspace{-0.1cm}
		\includegraphics[trim=1.cm 0.25cm 0.25cm 1.cm, clip=true, angle=0, scale=0.2625]{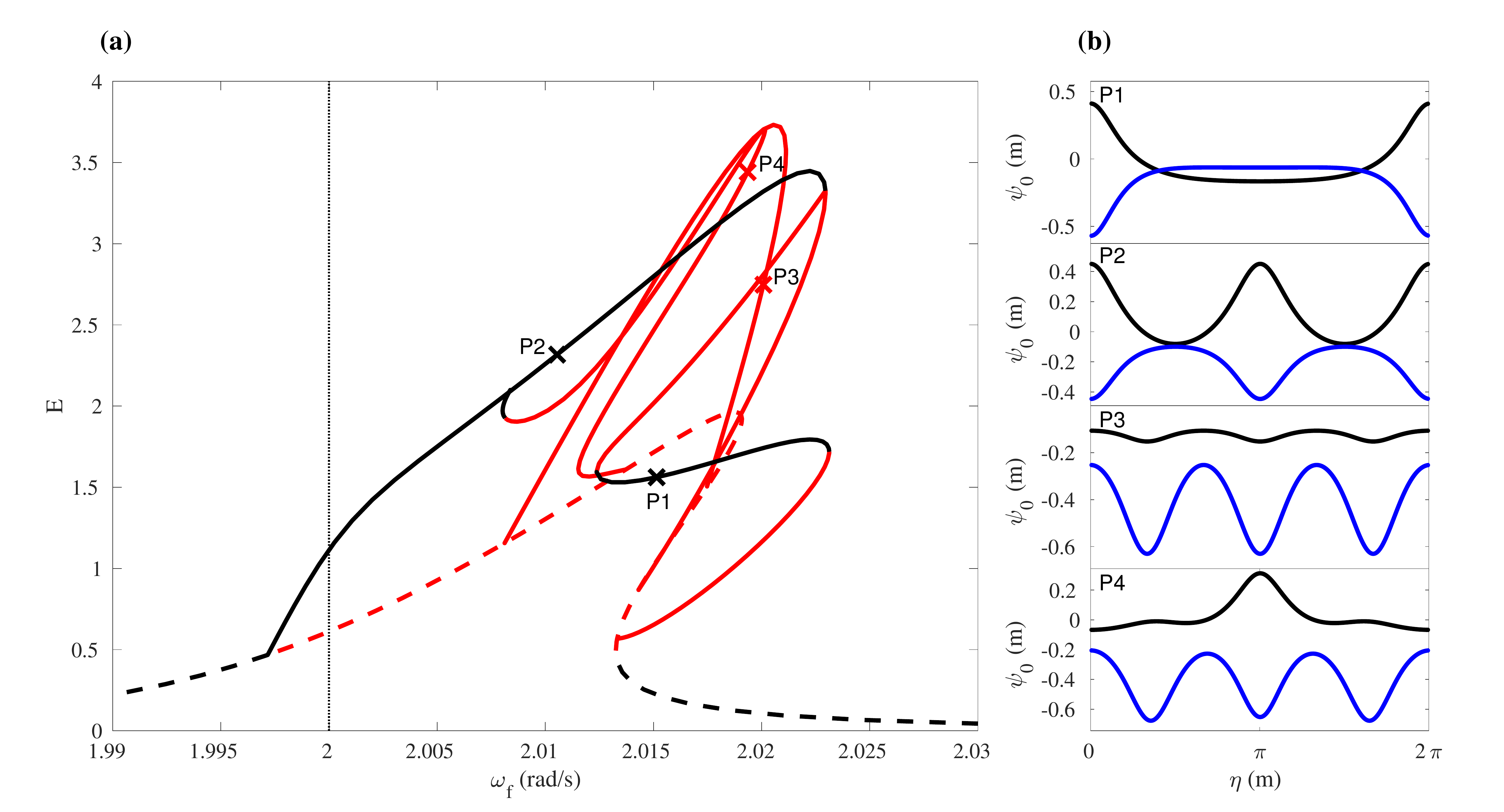}
		\caption{Solutions obtained from the NLSE implementation in Eq.~\eqref{Eq:FinalPsi} for $k$=16 m$^{-1}$ and $F_0$=0.010 ms$^{-2}$. Panels (a) and (b) illustrate the same quantities as in Fig. \ref{Fig:h1}.}
		\label{Fig:h2}
		\vspace{0.5cm}
		\includegraphics[trim=1.cm 0.25cm 0.25cm .cm, clip=true, angle=0, scale=0.2625]{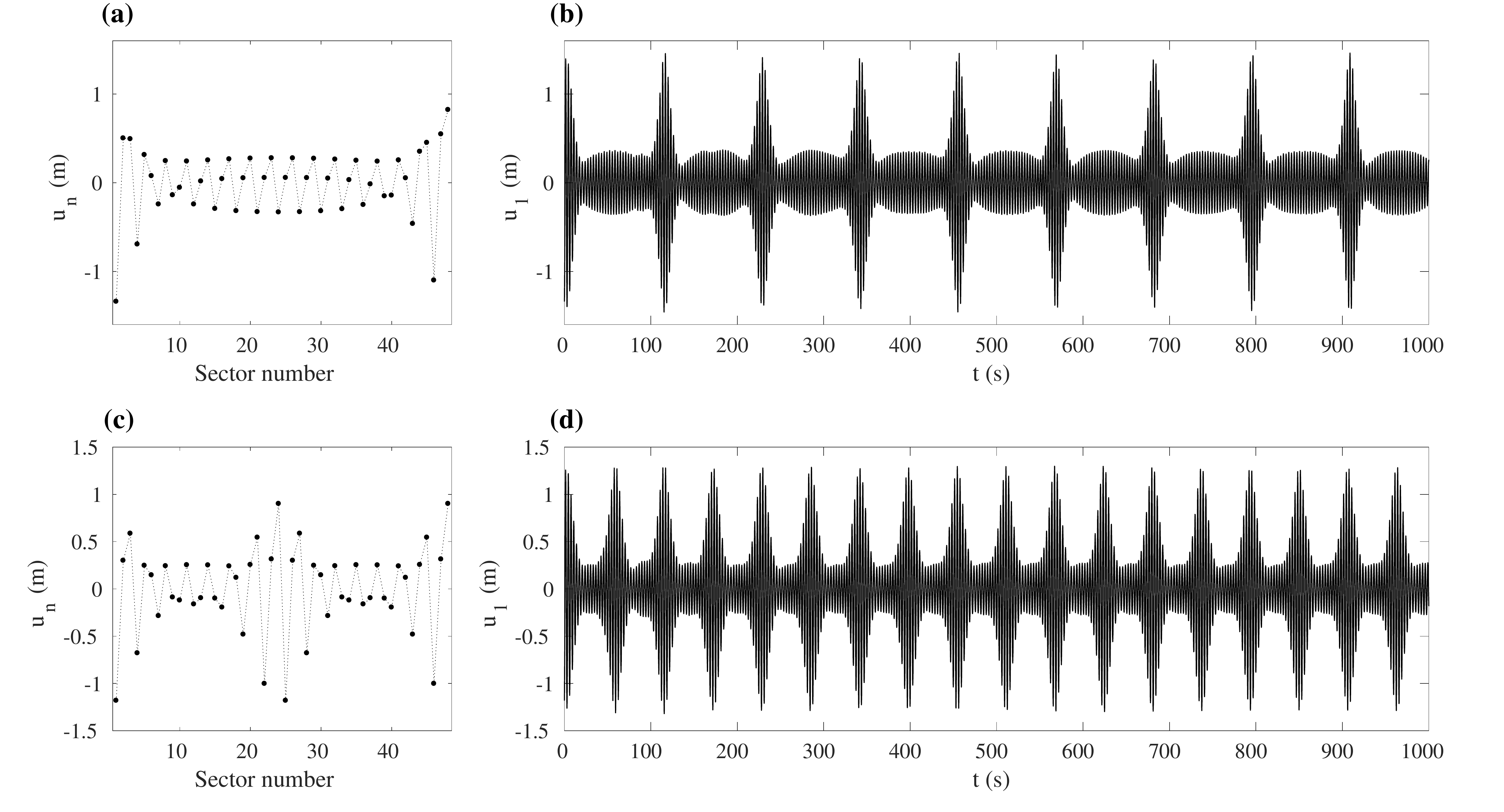}
		\caption{Results obtained through numerical integration of the physical system for $k$=16 m$^{-1}$ and $F_0$=0.010 ms$^{-2}$. Panels (a), (b), (c), and (d) illustrate the same quantities as in Fig. \ref{Fig:t_h1}.}
		\label{Fig:t_h2}
	\end{center}
\end{figure}

In the case of higher external forcing amplitude the bifurcation diagram is more complicated. Figure \ref{Fig:h2} displays results obtained in the same way as for Fig. \ref{Fig:h1}, but calculated assuming a stronger forcing of $F_0$=0.010 ms$^{-2}$ ($h$=0.0025 ms$^{-1}$). 
The result in Panel (a) shows that plane waves loose stability at lower values of $\omega_f$. Moreover, additional branches of localised solutions detach from the travelling waves. They are mainly composed of three humps solutions, as shown by P3 in Panel (b). However, most solutions are unstable, and the stable branches consist of one and two localised humps states.  

Solutions obtained from time integration of states P1 and P2, identified in Fig. \ref{Fig:h2}, are displayed in Fig. \ref{Fig:t_h2}. The results are similar to the ones presented in Fig. \ref{Fig:t_h1}, where solutions of one and two localised humps keep their shapes while they move around the system with approximately the group velocity. 

\begin{figure}[]
	\begin{center}
		\vspace{-0.1cm}
		\includegraphics[trim=1.cm 0.25cm 0.25cm 1.cm, clip=true, angle=0, scale=0.2625]{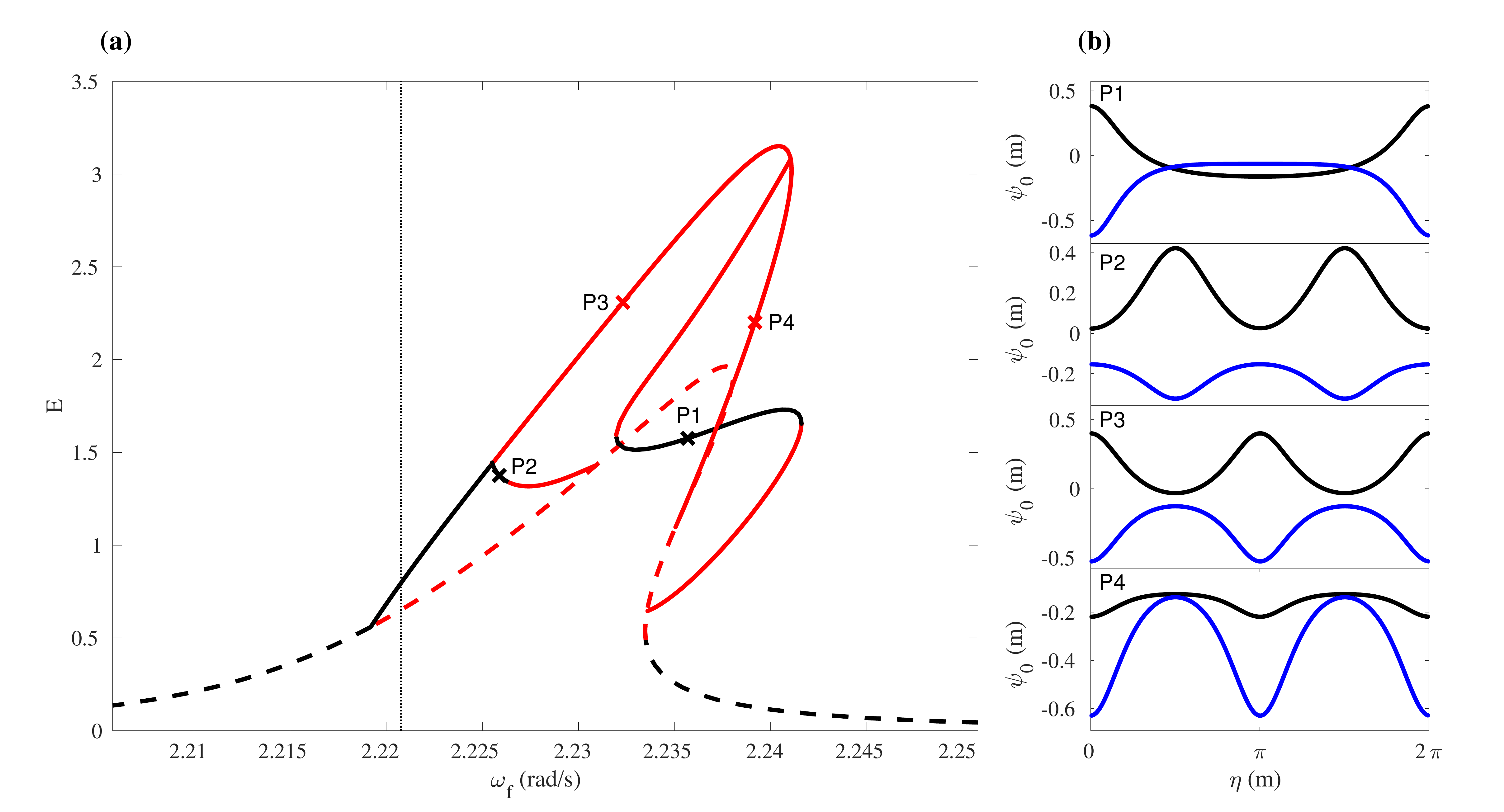}
		\caption{Solution obtained from the NLSE implementation in Eq.~\eqref{Eq:FinalPsi} for $k$=22 m$^{-1}$ and $F_0$=0.0011 ms$^{-2}$. Panels (a) and (b) illustrate the same quantities as in Fig. \ref{Fig:h1}.}
		\label{Fig:l1}
		\vspace{0.5cm}
		\includegraphics[trim=1.cm 0.25cm 0.25cm .cm, clip=true, angle=0, scale=0.2625]{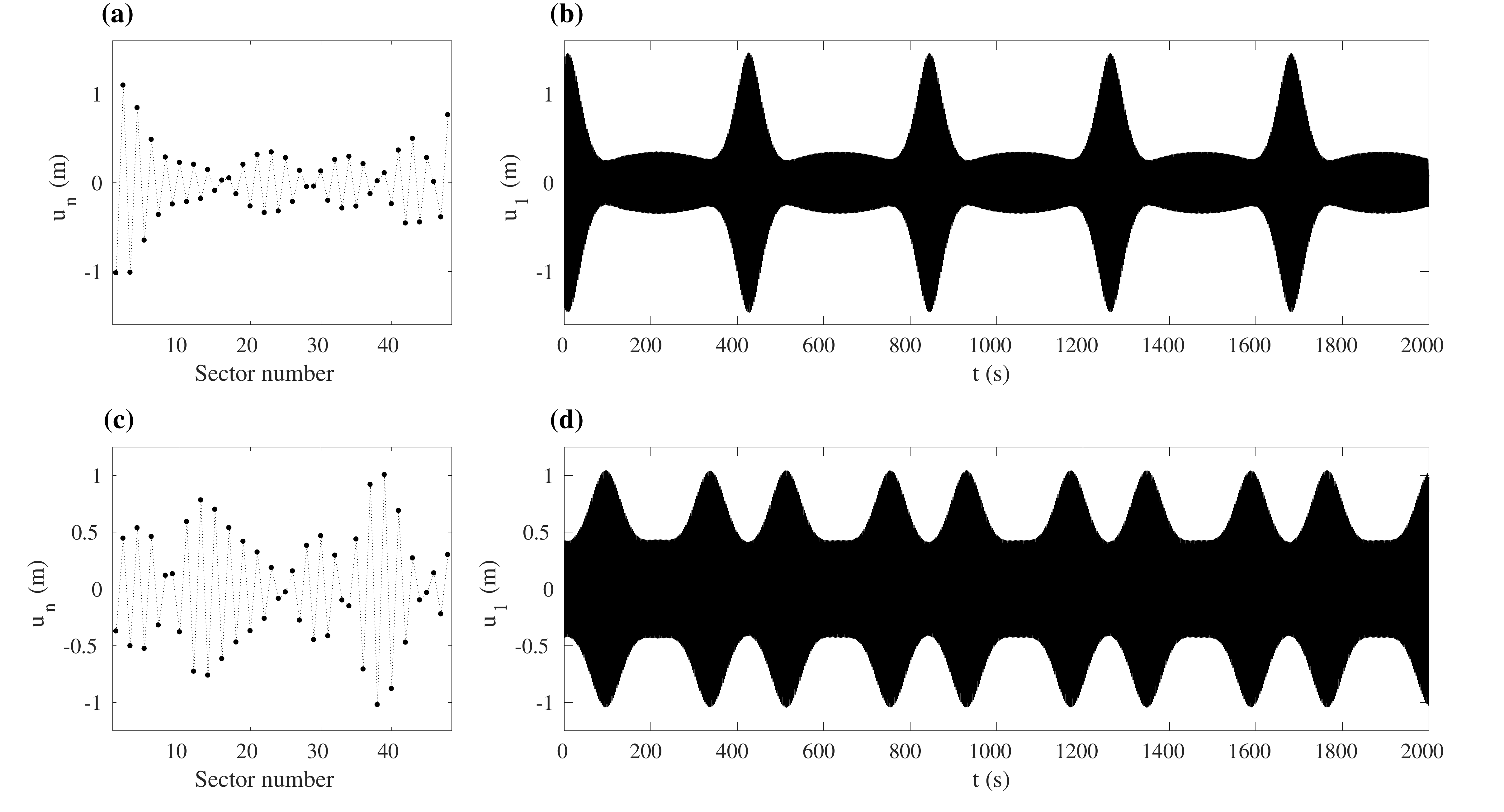}
		\caption{Results obtained through numerical integration of the physical system for $k$=22 m$^{-1}$ and $F_0$=0.0011 ms$^{-2}$. Panels (a), (b), (c), and (d) illustrate the same quantities as in Fig. \ref{Fig:t_h1}.}
		\label{Fig:t_l1}
	\end{center}
\end{figure}

In order to check the applicability of the NLSE approach to a different wave number, the same strategy is applied assuming $k$=22 m$^{-1}$ and $F_0$=0.0011 ms$^{-2}$. This configuration leads the parameters of Eq.~\eqref{Eq:FinalPsi} to $P$=-0.00378 m$^{2}$rad s$^{-1}$, Q=-0.00675 m$^{-2}$rad$^{-1}$s$^{-1}$, $\Gamma$=0.005 s$^{-1}$ and $h$=0.0025 ms${-1}$. The corresponding results are displayed in Fig. \ref{Fig:l1}. 

The results in Panel (a) and (b) are very similar to the ones already discussed in Figs. \ref{Fig:h1} and \ref{Fig:h2}. However, no stable solution is observed between $\omega_f \approx $ 2.226 rad/s and $\omega_f \approx$ 2.231 rad/s. This result suggests that stable configurations within this frequency range might not be symmetric with respect to $\eta=\pi$ and, therefore, we can not compute them with the present method.  

Solutions from points P1 and P2 are, again, used as initial conditions for time-integration of the physical system. The corresponding solutions are displayed in Fig. \ref{Fig:t_l1}. It is interesting to notice that, although solutions are symmetric with respect to $\eta=\pi$ m, the humps are not necessarily equidistant in space. This feature can be verified from $x_1(t)$ in Panel (d) of Fig. \ref{Fig:t_l1}, where the two humps are moving with the same velocity but they are clearly not equally spread over the physical system.

\begin{figure}[]
	\begin{center}
		\vspace{-0.1cm}
		\includegraphics[trim=1.cm 0.25cm 0.25cm 1.cm, clip=true, angle=0, scale=0.2625]{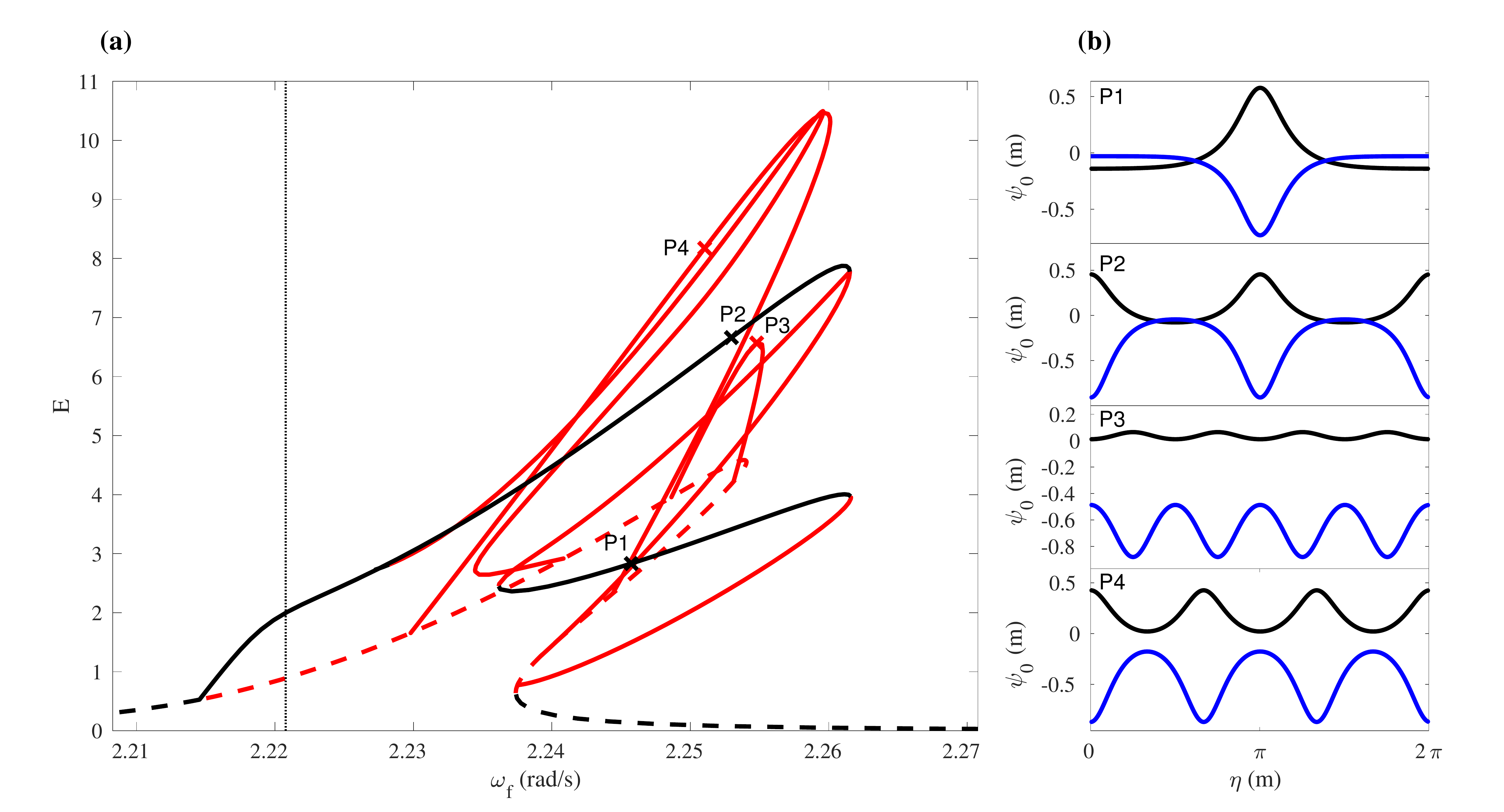}
		\caption{Solution obtained from the NLSE implementation in Eq.~\eqref{Eq:FinalPsi} for $k$=22 m$^{-1}$ and $F_0$=0.016 ms$^{-2}$. Panels (a) and (b) illustrate the same quantities as in Fig. \ref{Fig:h1}.}
		\label{Fig:l2}
		\vspace{0.5cm}
		\includegraphics[trim=1.cm 0.25cm 0.25cm .cm, clip=true, angle=0, scale=0.2625]{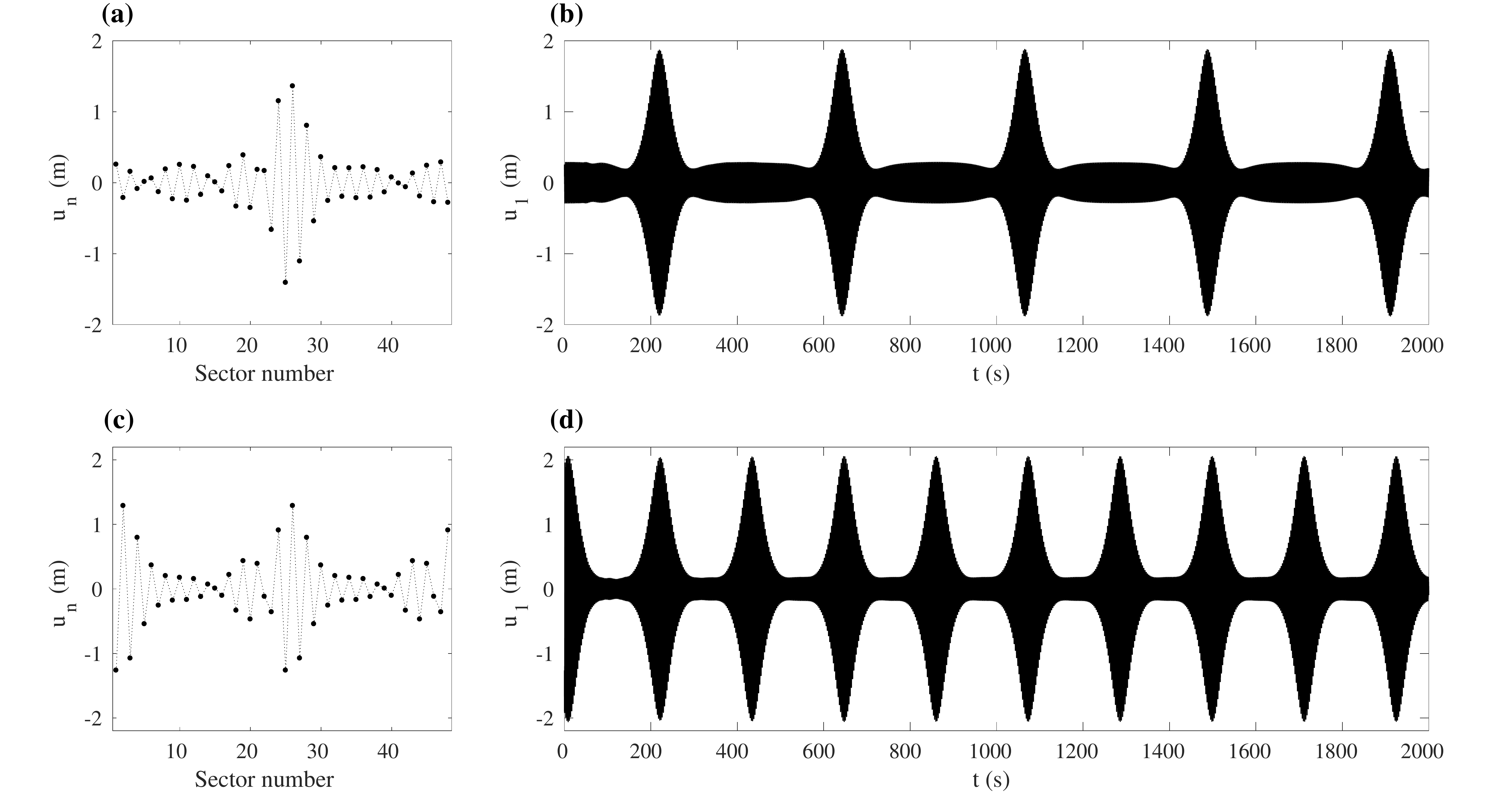}
		\caption{Results obtained through numerical integration of the physical system for $k$=22 m$^{-1}$ and $F_0$=0.016 ms$^{-2}$. Panels (a), (b), (c), and (d) illustrate the same quantities as in Fig. \ref{Fig:t_h1}.}
		\label{Fig:t_l2}
	\end{center}
\end{figure}

In the case of higher external forces amplitude, the results are very similar to the ones presented in Fig. \ref{Fig:h2}. Figure \ref{Fig:l2} displays the corresponding results calculated for $k$=22 m$^{-1}$ and $F_0$=0.016 ms$^{-2}$ ($h$=0.0035 ms$^{-1}$).

Again a quite complicated bifurcation diagram is observed. In this case, solutions with one, two, three and four humps detach from travelling waves. However, as already observed in the previous cases, most configurations are unstable. The stable solutions consist, again, of one (P1) and two (P2) hump solitons. 

Time-marching results in Fig. \ref{Fig:t_l2}, obtained from the equations of the physical system, show similar results as in Figs. \ref{Fig:t_h1}, \ref{Fig:t_h2}, and \ref{Fig:t_l1}. Localised states which keep their profile move around the system with roughly the group velocity.

\section{Continuation to the conservative limit} \label{Sec:Cons}

\begin{figure}[]
	\begin{center}
		\vspace{-0.1cm}
		\includegraphics[trim=1.cm 0.25cm 0.25cm 1.cm, clip=true, angle=0, scale=0.2625]{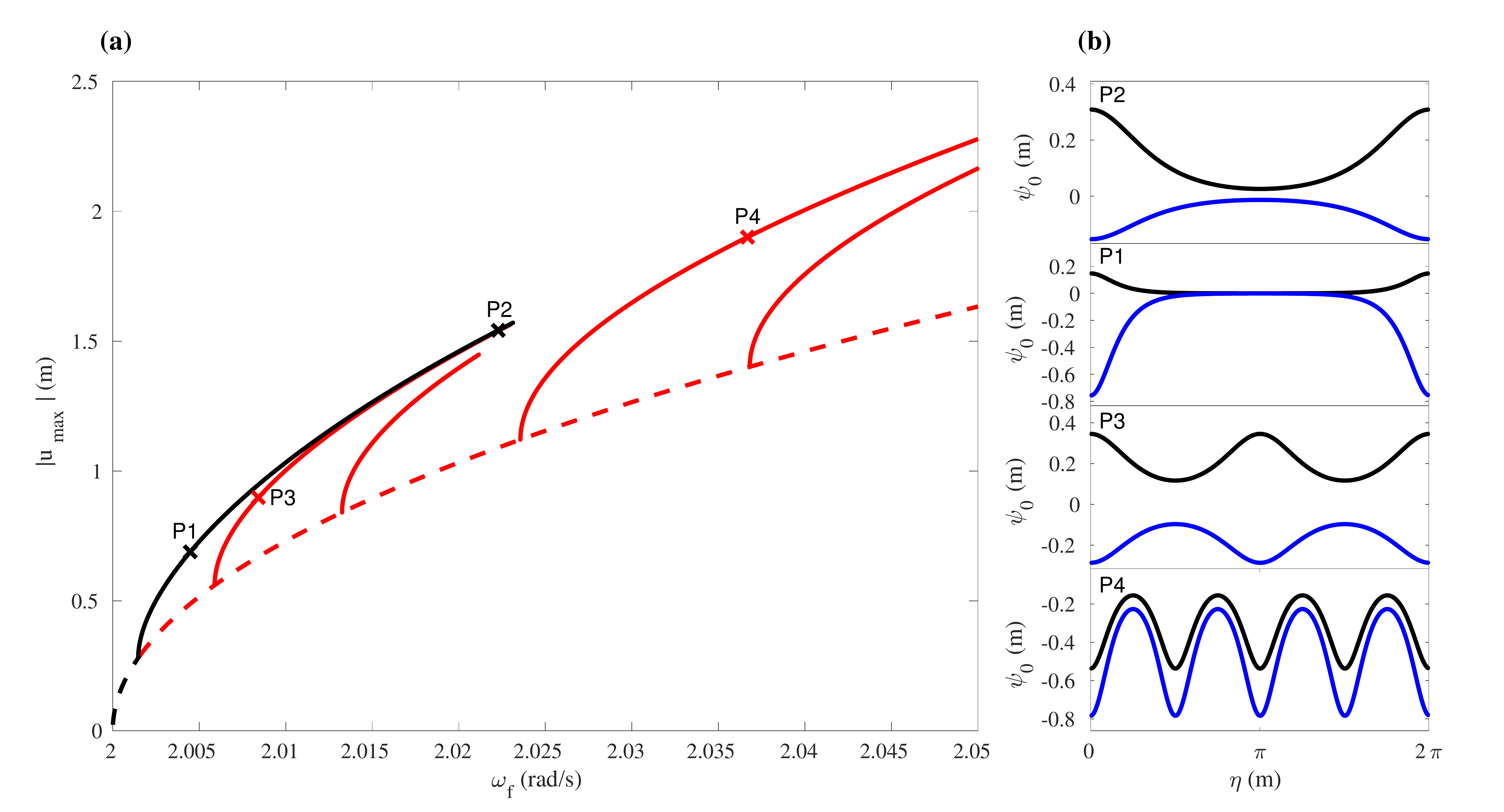}
		\caption{Localised states calculated for the conservative physical system for $k$=16 m$^{-1}$. Panel (a) displays the bifurcation diagram, where the black dashed line represents stable homogeneous solutions, while the red one shows the unstable regime. Full lines, detaching from travelling waves, represent stable (black) and unstable (red) localised solutions. Panel (b) shows the real (black) and imaginary (blue) parts of the $\psi_0$ solutions at P1, P2, P3, and P4 of Panel (a).}
		\label{Fig:cons1}
		\vspace{0.5cm}
		\includegraphics[trim=1.cm 0.25cm 0.25cm .cm, clip=true, angle=0, scale=0.2625]{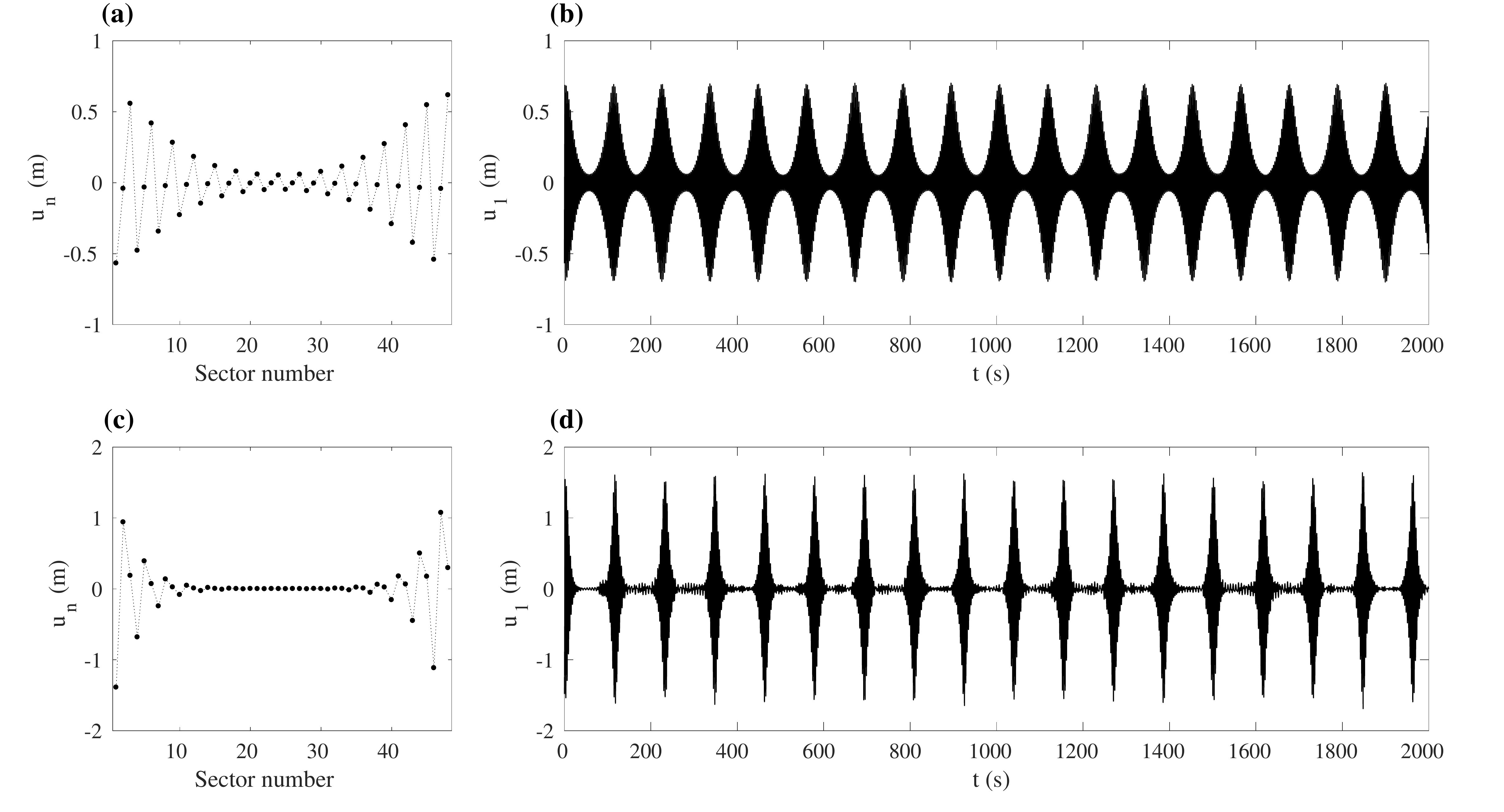}
		\caption{Results obtained through numerical integration of the conservative physical system $k$=16 m$^{-1}$. Panels (a), (b), (c), and (d) illustrate the same quantities as in Fig. \ref{Fig:t_h1}.}
		\label{Fig:t_cons1}
	\end{center}
\end{figure}

The analysis in Sec. \ref{Sec:NR} demonstrated that localised states branch off from unstable travelling waves. In order to understand how these dissipative solitons are related to the solitons in the conservative limit, a continuation from the driven-dissipative to the conservative case is conducted. To link the regimes, the autonomous NLSE in Eq.~\eqref{Eq:FinalPsi} is rewritten as 
\begin{equation}
\mbox{i} \frac{\partial \psi}{\partial \tau} - \delta_\omega \psi + P \frac{\partial^2 \psi}{\partial \eta^2} + Q |\psi|^2\psi =  - \zeta(\mbox{i} \Gamma \psi + h),
\label{Eq:Cons}
\end{equation}
where $\zeta$ is a continuation parameter. One should note that continuing $\zeta$ from $\zeta$=1 to $\zeta$=0 allows to follow driven-dissipative solutions of Sec. \ref{Sec:NR} into the conservative regime. Figure \ref{Fig:cons1} displays the results obtained for the wave number $k$=16 m$^{-1}$. 
In this case, contrary to analytical solutions of the NLSE in an infinite spatial domain, travelling waves lose stability at non-zero amplitudes. Moreover, at this point, a branch of stable localised solution composed of solitons with one hump detaches from the travelling waves. Initially, the soliton consists of a small hump on a travelling wave background. However, when the amplitude of the solution gets larger and nonlinear effects are stronger, this hump tends to localise more. Another important feature observed is that, at higher background amplitudes, other branches of localised states with two, three, four, and five humps detach, successively, from the homogeneous solutions. However, all the computed multi-solitons states are unstable within the dynamic analysed regime. 

In order to check the NLSE results, initial conditions resulting from points P1 and P2 in Fig. \ref{Fig:cons1} were computed to solve the equations of the physical system by time integration. Figure \ref{Fig:t_cons1} displays the corresponding results. It is possible to observe, in Panel (a) of Fig. \ref{Fig:t_cons1}, that the initial conditions consist of one hump modulating the travelling wave. Time-marching results, in Panel (b), show that this hump moves around the system keeping its shape, as expected from the NLSE theory. In Panel (c), where displacements are larger and the system is in a stronger non-linear regime, only a small part of the system vibrates, while the other part stands at rest. Results from time-integration, in Panel (d), show that in this case a zone of localised vibration moves around the system almost preserving its shape. However, it is possible to observe small vibrations in between the successive humps moving around the structure. These small vibrations are not expected from the NLSE results (see P2 in Fig. \ref{Fig:cons1}). This discrepancy is mainly due to the two assumptions imposed by the NLSE approximation: (1) the continuous approach, which looses accuracy when solutions are very localised; and (2) due to large displacements, when the single harmonic approximation is not sufficient anymore.

\begin{figure}[]
	\begin{center}
		\vspace{-0.1cm}
		\includegraphics[trim=1.cm 0.25cm 0.25cm 1.cm, clip=true, angle=0, scale=0.2625]{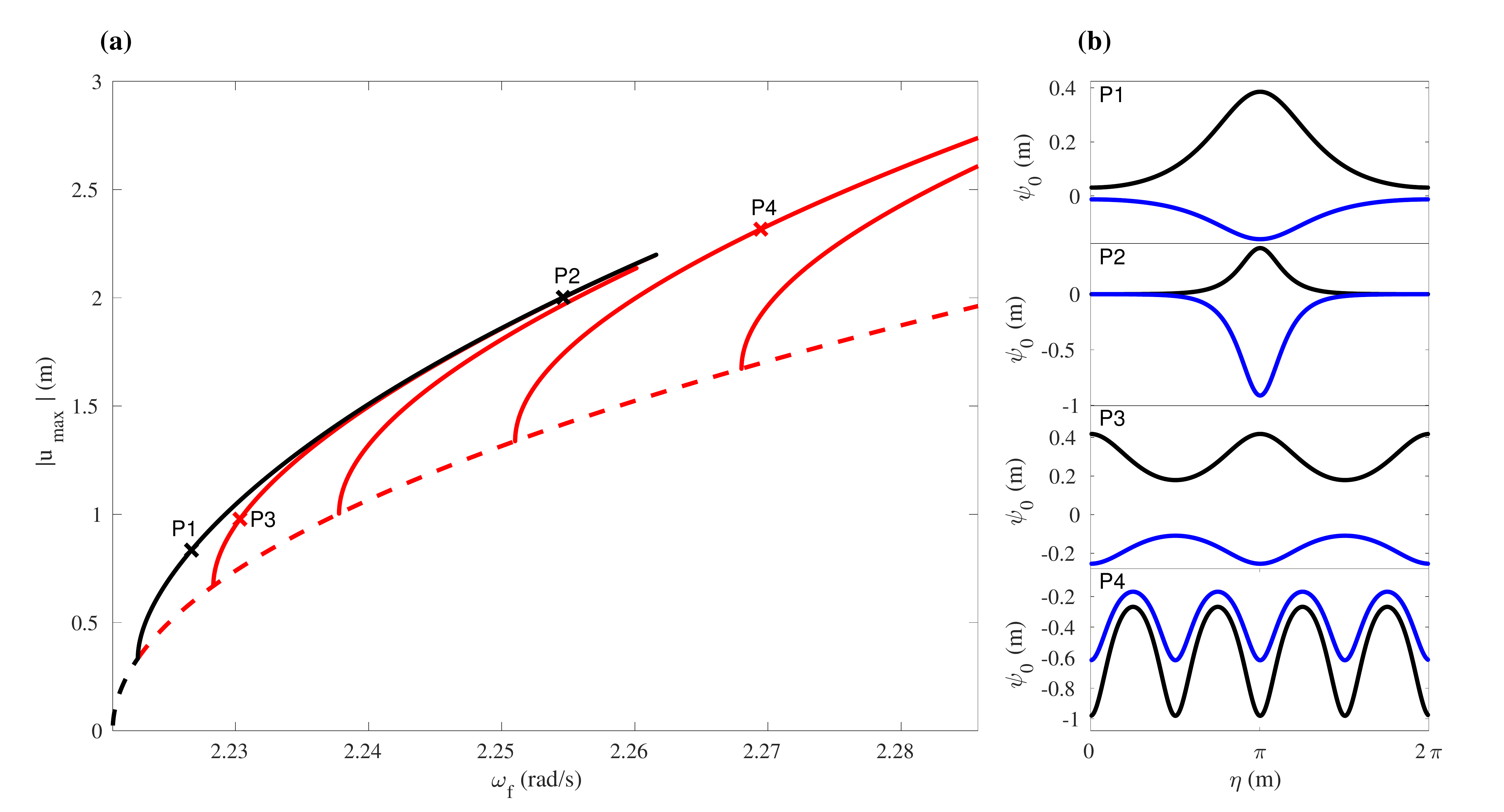}
		\caption{Results obtained for the conservative system assuming $k$=22 m$^{-1}$. Panels (a) and (b) illustrate the same quantities as in Fig. \ref{Fig:cons1}.}
		\label{Fig:cons2}
		\vspace{0.5cm}
		\includegraphics[trim=1.cm 0.25cm 0.25cm .cm, clip=true, angle=0, scale=0.2625]{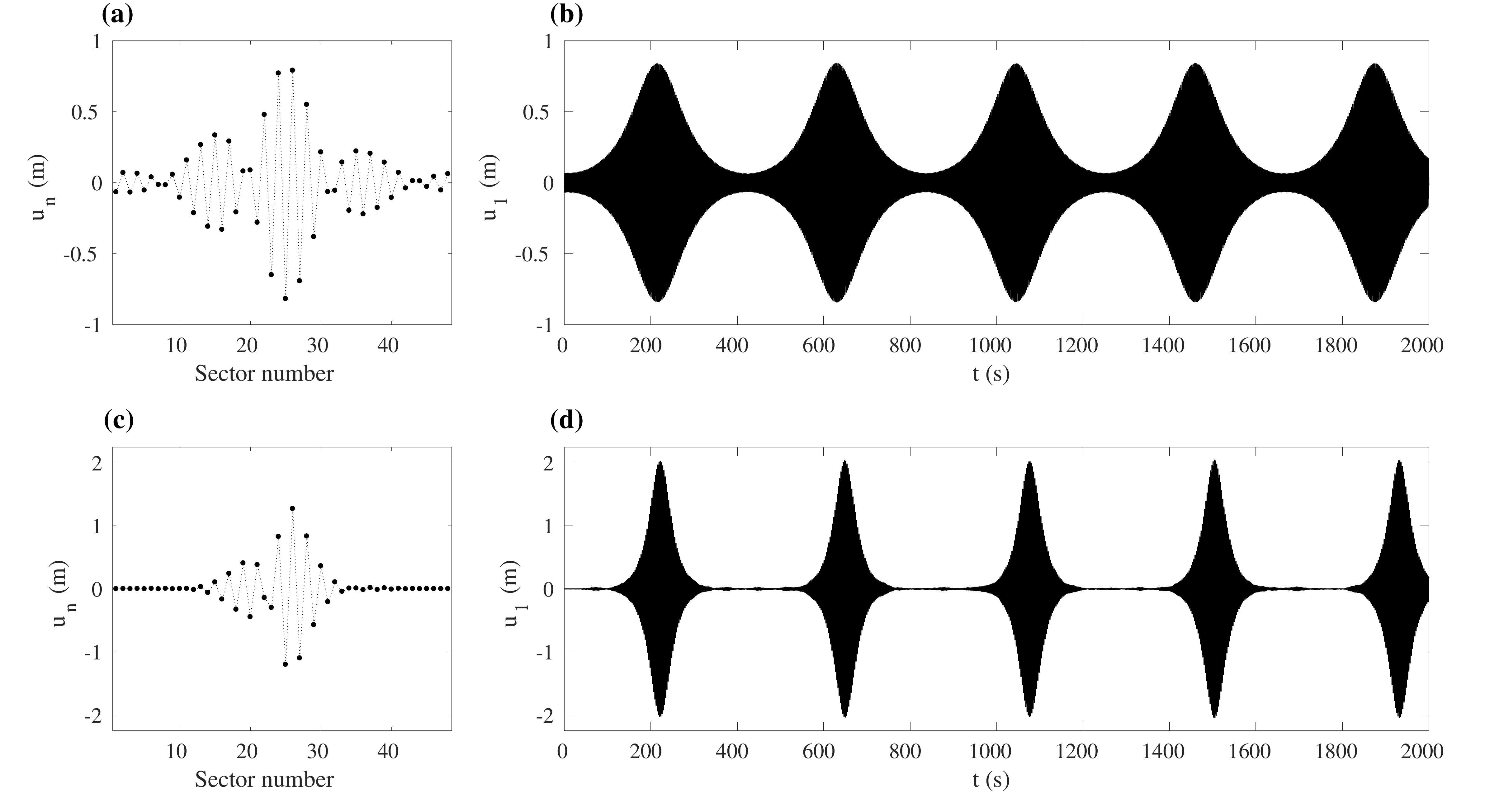}
		\caption{Results obtained through numerical integration of the conservative physical system for $k$=22 m$^{-1}$. Panels (a), (b), (c), and (d) illustrate the same quantities as in Fig. \ref{Fig:t_h1}.}
		\label{Fig:t_cons2}
	\end{center}
\end{figure}

Figure \ref{Fig:cons2} shows analogous results for another wave number $k$=22 m$^{-1}$.  
The solutions show, again, a one hump stable soliton detaching from unstable travelling waves with a non-zero amplitude. Other localised states with two, three, four, and five humps also detach from the homogeneous solution. However, as in the previous case, these multi-solitons solutions are unstable within the regime analysed.

Figure \ref{Fig:t_cons2} shows the corresponding results for P1 and P2 identified in Fig. \ref{Fig:cons2}. The results show, again, that a localised hump moves around the system preserving its shape.

\section{Conclusions and outlook} \label{Sec:Conc}

This paper focuses on localised oscillations arising from modulation of nonlinear travelling waves in weakly nonlinear cyclic and symmetric structures. The study is based on a minimal model composed of Duffing oscillators in the presence of linear viscous dampers and external travelling wave forces. The system can be understood as a minimal model of a bladed disk operating in the large deformations regime. It is demonstrated that, in the non-conservative regime, localised states composed of single and multi-soliton solutions bifurcate from travelling waves. Moreover, when these localised states are continued to the conservative regime, it is shown that solitons branch off from plane waves at finite amplitudes. 

In future work, more complex physical models need to be studied. Next investigations will focus on more complicated models, e.g. with two degrees of freedom per sector which also include the disk inertia. More realistic models, e.g. based on reduced-order systems obtained from full finite element models, should also be in the focus of future investigations. The phenomenological findings discussed in this paper also need to be compared to small-scale experiments. In addition, localised vibrations arising from other nonlinear models, such as friction and impact, are also an avenue of future research.

\section*{Acknowledgements}
	The authors thank Dr. Igor Barashenkov for discussions and advice regarding the non-conservative NLSE. F. Fontanela is funded by the Brazilian	National Council for the Development of	Science	and	Technology (CNPq) under the grant 01339/2015-3. The authors thank Rolls-Royce plc for allowing this publication. 
\bibliography{mybibfile3}

\appendix

\section{Derivation of the non-conservative NLSE}

The weakly nonlinear regime of the system described by Eq.~\eqref{Eq:Mot} is considered assuming a travelling wave excitation such as
\begin{eqnarray}
g_n(t)=F_0\exp\{\mbox{i}[k(n-1)a - \omega_k t] \}\exp\{\mbox{i}\delta_\omega t\} + \mbox{c.c.},
\end{eqnarray}
where the external force frequency $\omega_f$ is decomposed using the linear natural frequency $\omega_k$ and a detuning parameter $\delta_\omega$ such as $\omega_F=\omega_k - \delta_\omega$.
In order to derive a NLSE in the presence of linear viscous damping and external travelling wave excitation, it is assumed that the system is in the weakly nonlinear regime and experiencing travelling waves modulated by an envelope function such as
\begin{equation}
u_n=\epsilon \Psi(X,T) \exp\{\mbox{i}[k(n-1)a - \omega_k t] \} + \mbox{c.c.},
\end{equation}
where $X=\epsilon x$ and $T=\epsilon t$ are two slowly-varying functions. Therefore, the corresponding velocity $\dot{u}_n$ and acceleration $\ddot{u}_n$ of the $n$th mass are written as 
\begin{eqnarray}
\dot{u}_n = \left( \epsilon^2 \frac{\partial{\Psi}}{\partial T} - \mbox{i}\omega_k\epsilon\Psi\right) \exp\{\mbox{i}[k(n-1)a - \omega_k t] \} +  \mbox{c.c.}, \\
\ddot{u}_n = \left( \epsilon^3 \frac{\partial^2{\Psi}}{\partial T^2} - 2\mbox{i}\omega_k\epsilon^2\frac{\partial{\Psi}}{\partial T}- \omega_k^2\epsilon \Psi\right) \exp\{\mbox{i}[k(n-1)a - \omega_k t] \} +  \mbox{c.c.}. 
\end{eqnarray}

In the following, in order to introduce a continuum approximation, the envelope function at a neighbour degree of freedom $\Psi(na \pm a)$ is computed using a Taylor expansion such as
\begin{equation}
\Psi(na \pm a)=\Psi \pm a\epsilon \frac{\partial \Psi}{\partial X} +\frac{a^2\epsilon^2}{2} \frac{\partial^2 \Psi}{\partial X^2} + o(> \epsilon^2),
\end{equation}
where $o(> \epsilon^2)$ stands for terms higher than $\epsilon^2$. Moreover, the following additional assumptions are considered: (1) the damping $\gamma^2$ and the detuning parameter $\delta_\omega$ are at order $\epsilon^2$; and (2) the external force is at order $\epsilon^3$. After substituting the previous expressions into Eq.~\eqref{Eq:Mot}, ignoring higher-order harmonics, and considering the dispersion relation $\omega_k^2=\omega_0^2 + 4\omega_c^2 \sin^2 \left( \frac{ka}{2}\right)$, the process leads to 
\begin{multline}
\mbox{i} \epsilon^2 \left\lbrace  \frac{\partial \Psi}{\partial T} + c_g \frac{\partial \Psi}{\partial X} \right\rbrace + \\ \frac{\epsilon^3}{2 \omega_k} \left\lbrace \Lambda \frac{\partial^2 \Psi}{\partial X^2} - \frac{\partial^2 \Psi}{\partial T^2} -3\xi |\Psi|^2\Psi + \mbox{i}\omega_k \gamma^2 \Psi + F_0\exp\{\mbox{i}\epsilon\delta_{\omega}T\} \right\rbrace=0,
\label{Eq:Sch1}
\end{multline}
where $c_g=\frac{2a\omega_c^2}{\omega_k} \sin(ka)$ is the group velocity and $\Lambda=\omega_c^2a^2\cos(ka)$. In order to simplify Eq.~\eqref{Eq:Sch1}, it is rewritten in a frame $\eta$ moving with $c_g$ such as $\eta=X-c_gT$ and $\tau=\epsilon T$. Therefore, the following relations are obtained 
\begin{eqnarray}
\frac{\partial \Psi}{\partial X}=\frac{\partial \Psi}{\partial \eta}, \label{Eq:Coord1}\\
\frac{\partial^2 \Psi}{\partial X^2}=\frac{\partial^2 \Psi}{\partial \eta^2}, \label{Eq:Coord2}\\
\frac{\partial \Psi}{\partial T}=-c_g \frac{\partial \Psi}{\partial \eta} + \epsilon \frac{\partial \Psi}{\partial \tau},\\
\frac{\partial^2 \Psi}{\partial T^2}=c_g^2 \frac{\partial^2 \Psi}{\partial \eta^2} - 2 c_g \epsilon \frac{\partial^2 \Psi}{\partial \tau \partial \eta} + \epsilon^2 \frac{\partial^2 \Psi}{\partial \tau^2} \label{Eq:Coord3}.
\end{eqnarray}
After substituting Eqs.~\eqref{Eq:Coord1}--\eqref{Eq:Coord3} into Eq.~\eqref{Eq:Sch1} and ignoring terms higher than $\epsilon^3$, the processes leads to
\begin{equation}
\mbox{i} \frac{\partial \Psi}{\partial \tau} + \left( \frac{\Lambda - c_g}{2\omega_k} \right)  \frac{\partial^2 \Psi }{\partial \eta^2}  - \left( \frac{3\xi}{2\omega_k}\right) |\Psi|^2\Psi= -\mbox{i}\left( \frac{\gamma^2}{2}\right)  \Psi - \left( \frac{F_0}{2\omega_k} \right)  \exp\{\mbox{i}\delta_\omega \tau\}.
\label{Eq:Sch2}
\end{equation}
Equation~\eqref{Eq:Sch2} is an externally forced and linearly damped NLSE, which can be written in the form of Eq.~\eqref{Eq:AcNLS} by setting $P=\frac{1}{2}\frac{\mbox{d}^2\omega_k}{\mbox{d}k^2}=\frac{1}{2\omega_k}(\Lambda - c_g^2)$, Q=$\frac{-3\xi}{2\omega_k}$, $\Gamma=\frac{\gamma^2}{2}$, and $h=\frac{F_0}{2\omega_k}$.
\end{document}